\pdfoutput=1
\RequirePackage{ifpdf}
\ifpdf 
\documentclass[pdftex]{sigma}
\else
\documentclass{sigma}
\fi

\numberwithin{equation}{section}

\newtheorem{Theorem}{Theorem}[section]
\newtheorem*{Theorem*}{Theorem}
\newtheorem{Corollary}[Theorem]{Corollary}
\newtheorem{Lemma}[Theorem]{Lemma}
\newtheorem{Proposition}[Theorem]{Proposition}
 { \theoremstyle{definition}
\newtheorem{Definition}[Theorem]{Definition}

\newtheorem{Example}[Theorem]{Example}
 }

\usepackage{mathrsfs}

\def\abs#1{\left\vert #1 \right\vert}
\def\allpoly{\mbox{$\re\langle X \rangle$}}
\def\allpolyell{\mbox{$\re^{\ell}\langle X \rangle$}}
\def\allpolyx0degn{\mbox{$P_n$}}

\def\allseries{\mbox{$\re\langle\langle X \rangle\rangle$}}

\def\allseriesell{\mbox{$\re^{\ell} \langle\langle X \rangle\rangle$}}

\def\allseriesmLC{\mbox{$\re^{m}_{LC}\langle\langle X \rangle\rangle$}}

\def\allseriesellLC{\mbox{$\re^{\ell}_{LC}\langle\langle X \rangle\rangle$}}

\def\charseries{{\rm char}}

\newcommand{\comment}[1]{} 

\def\dim{{\rm dim}}
\def\dist{{\rm dist}}

\def\Endallseries{{\rm End}(\allseries)}

\def\eqref#1{(\ref{#1})} 

\def\mbf#1{\hbox{\mathversion{bold}$#1$}} 

\def\nat{{\mathbb N}} 
\def\norm#1{\left\Vert#1\right\Vert}

\def\re{{\mathbb R}} 

\def\shuffle{{\scriptscriptstyle \;\sqcup \hspace*{-0.05cm}\sqcup\;}}

\def\spanset{{\rm span}}
\def\supp{{\rm supp}}

\def\begals{\[\begin{aligned}}
\def\endals{\end{aligned}\]}
\def\begce{\begin{center}}
\def\endce{\end{center}}
\def\begar{\begin{array}}
\def\endar{\end{array}}
\def\begeq{\begin{equation}}
\def\endeq{\end{equation}}
\def\begdi{\begin{displaymath}}
\def\enddi{\end{displaymath}}
\def\begdis{\begin{eqnarray*}}
\def\enddis{\end{eqnarray*}}
\def\begeqa{\begin{eqnarray}}
\def\endeqa{\end{eqnarray}}
\def\begdes{\begin{description}}
\def\enddes{\end{description}}
\def\begit{\begin{itemize}}
\def\endit{\end{itemize}}
\def\begen{\begin{enumerate}}
\def\enden{\end{enumerate}}
\def\beglar{\left[\begin{array}}
\def\endrar{\end{array}\right]}
\def\begle{\begin{Lemma}}
\def\endle{\end{Lemma}}
\def\begde{\begin{Definition}}
\def\endde{\end{Definition}}
\def\begth{\begin{Theorem}}
\def\endth{\end{Theorem}}
\def\begco{\begin{Corollary}}
\def\endco{\end{Corollary}}
\def\begprop{\begin{Proposition}}
\def\endprop{\end{Proposition}}
\def\begex{\begin{Example}}
\def\endex{\end{Example}}
\def\begtab{\begin{tabular}}
\def\endtab{\end{tabular}}
\def\rref#1{(\ref{#1})}

\def\shuff#1#2{\mathbin{
 \hbox{\vbox{\hbox{\vrule \hskip#2 \vrule height#1 width 0pt}\hrule}\vbox{\hbox{\vrule \hskip#2 \vrule height#1 width 0pt\vrule }\hrule}}}}
\def\shuffl{{\mathchoice{\shuff{5pt}{3.5pt}}{\shuff{5pt}{3.5pt}}{\shuff{3pt}{2.6pt}}{\shuff{3pt}{2.6pt}}}}
\def\shuffle{{\, \shuffl \,}}
\def\output{f} 
\def\allseriesY{\mbox{$\re\langle\langle Y\rangle\rangle$}}
\def\allseriesXxY{\mbox{$\re\langle\langle X\otimes Y\rangle\rangle$}}

\begin{document}
\allowdisplaybreaks

\newcommand{\arXivNumber}{2202.10170}

\renewcommand{\thefootnote}{}

\renewcommand{\PaperNumber}{082}

\FirstPageHeading

\ShortArticleName{Entropy of Generating Series for Nonlinear Input-Output Systems}

\ArticleName{Entropy of Generating Series for Nonlinear\\ Input-Output Systems and Their Interconnections\footnote{This paper is a~contribution to the Special Issue on Non-Commutative Algebra, Probability and Analysis in Action. The~full collection is available at \href{https://www.emis.de/journals/SIGMA/non-commutative-probability.html}{https://www.emis.de/journals/SIGMA/non-commutative-probability.html}}}

\Author{W.~Steven GRAY}

\AuthorNameForHeading{W.S.~Gray}

\Address{Department of Electrical and Computer Engineering, Old Dominion University,\\ Norfolk, Virginia 23529, USA}
\Email{\href{mailto:sgray@odu.edu}{sgray@odu.edu}}
\URLaddress{\url{http://www.ece.odu.edu/~sgray/}}

\ArticleDates{Received February 22, 2022, in final form October 07, 2022; Published online October 25, 2022}

\Abstract{This paper has two main objectives. The first is to introduce a notion of entropy that is well suited for the analysis of nonlinear input-output systems that have a~Chen--Fliess series representation. The latter is defined in terms of its generating series over a noncommutative alphabet. The idea is to assign an entropy to a generating series as an element of a graded vector space. The second objective is to describe the entropy of generating series originating from interconnected systems of Chen--Fliess series that arise in the context of control theory. It is shown that one set of interconnections can never increase entropy as defined here, while a second set has the potential to do so. The paper concludes with a brief introduction to an entropy ultrametric space and some open questions.}


\Classification{68R15; 94A17; 93C10; 16T30}

\renewcommand{\thefootnote}{\arabic{footnote}}
\setcounter{footnote}{0}

\section{Introduction}

The concept of {\em entropy} in the context of dynamical systems has been defined in a variety of ways~\cite{Downarowicz_11,Young_03}.
Each notion is viewed as a measurement
of the rate of increase in dynamical complexity as the system evolves over time.
The idea has been adapted and applied in control theory starting with the work of Zames in the 1970s \cite{Zames_79}
in order to address engineering problems like controlled invariance, estimation, model detection,
and stabilization in a finite precision environment due to sampling, quantization, finite bit rates, etc.\ (see \cite{Liberzon_21} for a more complete overview).
More recently, the concept of entropy has been employed in the analysis of networked control systems \cite{Kawan-Delvenne_16,Matveev-etal_19,Savkin_06,Tomar-Zamani_20}.
The general goal there is to quantify the entropy of a network in terms of the entropies of its subsystems
in order to establish some fundamental performance bounds given limited communication between the subsystems.
In \cite{Liberzon_21}, for example,
an explicit bound on the topological entropy of a network of autonomous, continuous-time, nonlinear state space systems is derived in terms of
similar upper bounds on its subsystems. The main result is then applied to cascaded systems to produce an upper bound
on the entropy of the composite system.\looseness=-1

This paper has two main objectives.
The first is to introduce a notion of entropy that is well suited for the analysis of
input-output systems that have a Chen--Fliess series representation \cite{Fliess_81,Fliess_83}. The latter is
defined in terms of its generating series, i.e., a noncommutative formal power series over a finite alphabet.
The idea is to assign an entropy to a formal power series as an element of a graded vector space
which measures how quickly in an asymptotic sense the homogeneous components of the grading are being occupied by the support of the
series. The asymptotic behavior of the Hilbert--Poincar\'e series for the vector space provides a direct basis for
comparison \cite{Anick_82,Atiyah-MacDonald_69}.
This idea is closely related to the concept of entropy appearing in formal language theory using what is called
a {\em generating structure function} \cite{Kuich_70,Kuich-Maurer_71}. There the grading is always based on word length,
and the two notions of entropy coincide if the formal power series is taken to be the characteristic series of the language.
As the authors of these works explain, their definition is in turn related to the classical
information theoretic notion of entropy introduced by Shannon in terms of channel capacity \cite{Shannon_48}.
Other distinct definitions for the entropy of a language exist. For example, the concept introduced in \cite{Schneider-Borchmann_17} measures the complexity
of a language in terms of the exponential growth rate in the number of equivalence classes of the Nerode congruence relation of the language
as the word length increases. In this setting, all
rational languages and Dyck languages have zero entropy, which is not the case here or in earlier work.
The entropy of a formal power series as defined here is also related in spirit to the concept of entropy appearing in symbolic dynamics \cite{Lind-Marcus_21},
namely, the entropy defined for a semigroup under concatenation corresponding to the language of forbidden words in a shift space.
It is, however, distinct from this established definition in that there is no
underlying set of recurrence inequalities to provide
additional structure (see, for example, \cite[Lemma~4.1.7]{Lind-Marcus_21}). The proposed entropy concept
also coincides in certain special cases (modulo a~logarithm) with the notion of entropy defined for
graded algebras \cite{McLachlan-Ryland_03,Newman-etal_00}. But that connection does not appear to be so important for the present work.
Finally, it should be stated that the
entropy of a generating series can be used to develop notions of entropy for
its corresponding input-output map. One such example is
{\em system identification entropy} as defined in \cite{Gray_Allerton22}.
This quantity describes the growth in the number of bits needed to specify the series coefficients in order to approximate the
input-output map as the desired accuracy increases.

The second objective of the paper is to describe the entropy of generating series originating from interconnected systems
of Chen--Fliess series found in the context of control theory. Such interconnections have been studied extensively by the author
and others
\cite{Ferfera_79,Ferfera_80,Fliess_81,Foissy_15,Gray-etal_SCL14,Gray-Ebrahimi-Fard_SIAM17,Gray-Ebrahimi-Fard_SCL21, Gray-etal_SCL09,Gray-Li_SIAM05,Gray-Thitsa_IJC12,Gray-Venkatesh_SCL2019,Gray-Wang_MTNS08, Thitsa-Gray_12,Venkatesh_21,Venkatesh-Gray_21,Winter-Arboleda-etal_15}.
It is known that in every case the composite system always has a Chen--Fliess series
representation whose generating series can be computed in terms of an induced formal power series product
applied to the generating series of the component systems. The main products of interest are addition, the shuffle product,
the composition product and the feedback product. The Cauchy product and Hadamard product play a
supporting role in this setting.
The main question to be addressed here is which system interconnections/formal power series products
are capable of increasing entropy and which ones are not.
It will turn out that all linear systems have generating series with {\em zero} entropy,
so any interconnection that preserves linearity will not increase entropy.
Somewhat surprising, even interconnections like the parallel product of two linear systems, which do not
preserve linearity, will still result in a new system having zero entropy.
Thus, this new concept of entropy is nontrivial only for nonlinear systems.
Finally, this work in many ways is the {\em complement} of previous work by the author and others
regarding the convergence of Chen--Fliess series and their interconnections
\cite{Gray-Li_SIAM05,Gray-Thitsa_IJC12,Gray-Wang_02,Thitsa-Gray_12,Venkatesh_21,Venkatesh-Gray_21,Wang_90,Winter-Arboleda-etal_15}.
The main focus there was on the growth rate of the coefficients of a~generating series
and the specific types of convergence that are guaranteed for the corresponding Chen--Fliess series.
In this paper, the nature of the coefficients is almost entirely irrelevant aside from being either zero or nonzero,
i.e., being in the support of the generating series or not.
Nevertheless, as shown in~\cite{Gray_Allerton22},
the entropy of a generating series can be used to refine the notion of radius of convergence of a Chen--Fliess series as defined
in~\cite{Thitsa-Gray_12}.

The paper concludes with a brief introduction to an entropy ultrametric space. Such constructions have appeared in other
contexts, for example, \cite{Caracciolo-Radicati_89}. As feedback systems are often analyzed in an ultrametric space setting
\cite{Ferfera_79,Gray-etal_SCL14,Gray-Li_SIAM05,Gray-Wang_MTNS08},
this framework could have immediate applications in control theory.
Some current open problems are posed in this setting.

The paper is organized as follows. In the next section, a few preliminaries regarding
formal power series, Chen--Fliess series, and system interconnections are summarized.
In Section~\ref{sec:entropy-of-generating-series}, the definition of entropy of a
generating series is given along with a number of illustrative examples.
The entropy of interconnected systems is characterized in Section~\ref{sec:entropy-interconnections}.
The entropy ultrametric is introduced in the subsequent section.
The conclusions of the paper are summarized in the final section.

\section{Preliminaries}\label{sec:preliminaries}

An {\em alphabet} $X=\{ x_0,x_1,\dots,x_m\}$ is any nonempty and finite set
of symbols referred to as {\em
letters}. A {\em word} $\eta=x_{i_1}\cdots x_{i_k}$ is a finite sequence of letters from $X$.
The number of letters in a word $\eta$, written as $\abs{\eta}$, is called its {\em length}.
The empty word, $\varnothing$, is taken to have length zero.
For a given $x_i\in X$, $|\eta|_{x_i}$ is the number of times $x_i$ appears in $\eta$.
The collection of all words having length $k$ is denoted by
$X^k$. Define $X^+=\bigcup_{k\geq 1} X^k$ and $X^\ast=\bigcup_{k\geq 0} X^k$,
the latter of which is a noncommutative monoid under concatenation.
Any mapping $c\colon X^\ast\rightarrow
\re^\ell$ is called a {\em formal power series}.
It is often
written as the formal sum $c=\sum_{\eta\in X^\ast}( c,\eta)\eta$,
where the {\em coefficient} $(c,\eta)\in\re^\ell$ is the image of
$\eta\in X^\ast$ under $c$.
The {\em support} of $c$, $\supp(c)$, is the set of all words having nonzero coefficients.
The set of all noncommutative formal power series over the alphabet $X$ is
denoted by $\allseriesell$. The subset of series with finite support, i.e., polynomials,
is represented by $\allpolyell$.
As $\allpoly$ is dense in $\allseries$ (under the ultrametric topology \cite{Berstel-Reutenauer_88}),
$\allseriesell$ can be viewed as the completion of $\allpolyell$.
Given any language $L\subseteq X^\ast$, its {\em characteristic series} in $\allseries$ is $\charseries(L)=\sum_{\eta\in L}\eta$.

\subsection{Entropy of graded vector spaces}

Let $\nat=\{1,2,\dots\}$ and $\nat_0=\nat\cup\{0\}$.
An $\re$-vector space $V$ is said to be $\nat_0${\em-graded} over $\re$ if
$V=\bigoplus_{n\in\nat_0} V_n$, where each $V_n$ is an $\re$-vector subspace. Elements in $V_n$ are said to have
degree~$n$. $V$ is {\em connected} if $V_0=\re$ and {\em locally finite} if each $V_n$ has finite dimension.
The {\em Hilbert--Poincar\'e series} of a graded locally finite vector space is defined to be
the formal power series in indeterminate $z$~\cite{Anick_82,Atiyah-MacDonald_69}
\[
V(z)=\sum_{n=0}^\infty \dim(V_n)z^n.
\]
Often $V(z)$ is a rational function indicating some type of linear recursion relates the
sequence of dimensions $d_n:=\dim(V_n)$, $n\geq 0$.
The {\em entropy} of a graded vector space $V$ is taken to be
\[
H(V)=\limsup_{n\rightarrow \infty} \sqrt[n]{\dim(V_n)},
\]
whenever the limit exists \cite{Newman-etal_00}.

\begex \label{ex:graded-vector-space}
Consider $\allseries$ with scalar multiplication and addition defined in the usual way to
form an $\re$-vector space.
If $\allseries$ is graded by word length, then it is connected and locally finite with
$V_n=\spanset_\re\{\eta\in X^\ast\colon |\eta|=n\}$ and
$d_n=\dim(V_n)=(m+1)^n$ so that
\[
V(z)=\sum_{n=0}^\infty (m+1)^nz^n=\frac{1}{1-(m+1)z}.
\]
As the sequence of dimensions is a geometric sequence, clearly, $d_{n+1}=(m+1)d_n$, $n\geq 0$ with $d_0=1$.
The entropy is $H(\allseries)=m+1$.

Consider next an alternative grading of $\allseries$, where the letter $x_0$
has twice the degree of the other letters and $\deg(\varnothing)=1$, that is, $\deg(\eta)=2|\eta|_{x_0}+\sum_{i=1}^m |\eta|_{x_i}+1$ for all $\eta\in X^\ast$
\cite{Foissy_15,Gray-etal_SCL14}. Let $|A|$ denote the cardinality of set $A$.
Define the formal power series in commuting indeterminates $z_0,z_1,\dots z_m$
\begin{align*}
W(z_0,z_1,\dots,z_m)&=\sum_{k_0,k_1,\dots,k_m=0}^\infty |\{\eta\in X^\ast\colon |\eta|_{x_j}=k_j,
j=0,1,\dots,m\}|
z_0^{k_0}z_1^{k_1}\cdots z_m^{k_m} \\
&=\frac{1}{1-(z_0+z_1+\cdots +z_m)}.
\end{align*}
In light of the assumed grading, it follows that
\begin{gather*}
V(z)= 1+mzW\big(z^2,z,\dots,z\big) =\frac{1-z^2}{1-mz-z^2} \\
\hphantom{V(z)}{}
= 1+m z+m^2 z^2+\big(m+m^3\big) z^3+
\big(2 m^2+m^4\big) z^4+\big(m+3 m^3+m^5\big) z^5 \\
\hphantom{V(z)=}{}+\big(3 m^2+4 m^4+m^6\big) z^6+\big(m+6m^3+5m^5+m^7\big)z^7+O\big(z^8\big).
\end{gather*}
In which case, $d_0=1$, $d_1=m$, $d_2=m^2$,
\[
d_{n+1}=md_{n}+d_{n-1},\qquad n\geq 2
\]
and
\[
d_n\sim \frac{m}{\sqrt{m^2+4}}\left(\frac{m+\sqrt{m^2+4}}{2}\right)^n,\qquad n\gg 1.
\]
If $m=1$, then $d_n$, $n\geq 1$ is the Fibonacci sequence, and
$d_n\sim \varphi^n/\sqrt{5}$
with $\varphi=\big(1+\sqrt{5}\big)/2$ being the golden ratio. The integer sequences for $m=2$ and $m=3$ are also
well studied
(see A052542 and A052906, respectively, in \cite{OEIS}).
Under this alternative grading, the entropy is $H(\allseries)=\big(m+\sqrt{m^2+4}\big)/2<m+1$.
\endex

\subsection{Elementary products of formal power series}

Suppose multiplication on $\re^\ell$ is defined componentwise.
Then there are three elementary products on $\allseriesell$ that
render it an associative unital $\re$-algebra \cite{Fliess_74,Fliess_81}.
The first is the commutative {\em Hadamard product} defined as
\[
c\odot d=\sum_{\eta\in X^\ast} (c,\eta)(d,\eta) \eta.
\]
The second is the noncommutative {\em Cauchy product}
\[
cd=\sum_{\eta,\xi\in X^{\ast}} (c,\eta)(d,\xi)\,\eta\xi.
\]
Lastly, $\allseriesell$ is a commutative $\re$-algebra under the bilinear {\em shuffle product}
\[
c\shuffle d=\sum_{\eta,\xi\in X^{\ast}} (c,\eta)(d,\xi)\,\eta\shuffle\xi,
\]
where the shuffle product of two words $x_i\eta,x_j\xi\in X^\ast$ is defined inductively by
\[
	(x_i\eta)\shuffle(x_j\xi)=x_i(\eta\shuffle(x_j\xi))+x_j((x_i\eta)\shuffle \xi)
\]
with $x_i,x_j\in X$ and $\eta\shuffle\varnothing=\varnothing\shuffle\eta=\eta$.
Each product is locally finite. In addition,
\begin{gather*}
\supp(c+d)\subseteq \supp(c)\cup\supp(d), \\
\supp(c\odot d)=\supp(c)\cap\supp(d),\\
\supp(cd)\subseteq \supp(c)\supp(d).
\end{gather*}
The relationship between the support of the shuffle product and that of its arguments is
much more complicated, see, for example, \cite{Lothaire_97}. Enumerating the number of distinct words from shuffle products
was investigated in \cite{Smith_09}. But asymptotic results along these lines, which would be very useful here, appear to be
unknown at present.

\subsection{Chen--Fliess series}
\label{subsec:CF-series}

Given any $c\in\allseriesell$, one can associate a causal
$m$-input, $\ell$-output operator, $F_c$, in the following manner.
Let $\mathfrak{p}\ge 1$ and $t_0 < t_1$ be given. For a Lebesgue measurable
function $u\colon [t_0,t_1] \rightarrow\re^m$, define
$\norm{u}_{\mathfrak{p}}=\max\{\norm{u_i}_{\mathfrak{p}}\colon 1\le
i\le m\}$, where $\norm{u_i}_{\mathfrak{p}}$ is the usual
$L_{\mathfrak{p}}$-norm for a measurable real-valued function,
$u_i$, defined on $[t_0,t_1]$. Let $L^m_{\mathfrak{p}}[t_0,t_1]$
denote the set of all measurable functions defined on $[t_0,t_1]$
having a finite $\norm{\cdot}_{\mathfrak{p}}$ norm and
$B_{\mathfrak{p}}^m(R)[t_0,t_1]:=\{u\in
L_{\mathfrak{p}}^m[t_0,t_1]\colon \norm{u}_{\mathfrak{p}}\leq R\}$.
Assume $C[t_0,t_1]$
is the subset of continuous functions in $L_{1}^m[t_0,t_1]$. Define
inductively for each word $\eta=x_i\bar{\eta}\in X^{\ast}$ the map $E_\eta\colon
L_1^m[t_0, t_1]\rightarrow C[t_0, t_1]$ by setting
$E_\varnothing[u]=1$ and letting
\[E_{x_i\bar{\eta}}[u](t,t_0) =
\int_{t_0}^tu_{i}(\tau)E_{\bar{\eta}}[u](\tau,t_0)\,{\rm d}\tau, \] where
$x_i\in X$, $\bar{\eta}\in X^{\ast}$, and $u_0=1$. The
{\em Chen--Fliess series} corresponding to $c\in\allseriesell$ is~\cite{Fliess_81}
\begin{gather}
y(t)=F_c[u](t) =
\sum_{\eta\in X^{\ast}} ( c,\eta) E_\eta[u](t,t_0). \label{eq:Fliess-operator-defined}
\end{gather}
To establish the convergence of this series, assume
there exist real numbers $K,M>0$ such that
\begin{gather}
\abs{(c,\eta)}\le K M^{|\eta|}|\eta|!,\qquad \forall\eta\in X^{\ast}.\label{eq:local-convergence-growth-bound}
\end{gather}
(Define $\abs{z}:=\max_i \abs{z_i}$ whenever $z\in\re^\ell$.)
It is shown in \cite{Gray-Wang_02}
that under such circumstances
the series \rref{eq:Fliess-operator-defined} converges uniformly and
absolutely so that $F_c$ describes a well defined mapping from
$B_{\mathfrak p}^m(R)[t_0,$ $t_0+T]$ into $B_{\mathfrak
q}^{\ell}(S)[t_0, \, t_0+T]$ for sufficiently small $R,T>0$, where
$\mathfrak{p},\mathfrak{q}\in[1,\infty]$ satisfy $1/\mathfrak{p}+1/\mathfrak{q}=1$.
The operator $F_c$ is said to be {\em locally convergent} and
is called a {\em Fliess operator}.
The collection of all generating
series $c$ satisfying
the growth condition \rref{eq:local-convergence-growth-bound}
is denoted by $\allseriesellLC$.

A special class of Chen--Fliess series consists of those having an input-output map $y=F_c[u]$ with
a smooth $n$ dimensional state space realization
\[
\dot{z}=g_0(z)+ \sum_{i=1}^mg_i(z)u_i,\qquad z(0)=z_0,\qquad y=\output(z)
\]
on some open set $W\subseteq\re^n$.
In this case, the generating series is computed by
\begin{gather} \label{eq:FPS-differential-representation}
(c,\eta)=L_{\eta}\output(z_0):=L_{g_{i_1}}\cdots L_{g_{i_k}}\output(z_0)
\end{gather}
for any $\eta=x_{i_k}\cdots x_{i_1}\in X^\ast$,
where $z_0\in\re^n$, and $L_{g_i}\output$ is the Lie derivative of the smooth output function $\output\colon W\rightarrow\re$ with respect to
vector field $g_i$ \cite{Fliess_81,Isidori_95}.
The {\em control Lie algebra}, $\mathscr C(g)$, for this realization is the smallest Lie algebra
containing $g_i$, $i=0,1,\dots,m$ which is closed under the Lie bracket operation
\[
[g_i,g_j](z)=\frac{\partial g_j}{\partial z} g_i(z)-\frac{\partial g_i}{\partial z} g_j(z).
\]
Identifying $g_i\sim x_i$ and letting $[x_i,x_j]=x_jx_i-x_ix_j$, ${\mathscr C}(g)$ is isomorphic to a Lie subalgebra ${\mathscr C}(X)$
of the free Lie algebra over~$X$.
${\mathscr C}(X)$ will be referred to as the control Lie algebra over~$X$.

\subsection{System interconnections}\label{subsec:system-interconnections}

Given Fliess operators $F_c$ and $F_d$, where $c,d\in\allseriesellLC$,
the parallel sum and parallel product connections satisfy $F_c+F_d=F_{c+d}$ and $F_cF_d=F_{c\shuffle d}$,
respectively~\cite{Fliess_81}.
When Fliess operators~$F_c$ and~$F_d$ with $c\in\allseriesellLC$ and
$d\in\allseriesmLC$ are interconnected in a cascade fashion, the composite
system $F_c\circ F_d$ has the
Fliess operator representation $F_{c\circ d}$, where
the {\em composition product} of~$c$ and~$d$
is given by~\cite{Ferfera_79,Ferfera_80}
\[
c\circ d=\sum_{\eta\in X^\ast} ( c,\eta) \psi_d(\eta)(\mathbf{1}).
\]
Here $\mbf{1}$ denotes the monomial $1\varnothing$, and
$\psi_d$ is the continuous (in the ultrametric sense)
algebra homomorphism
from $\allseries$ to the set $\Endallseries$ of vector space endomorphisms on $\allseries$ uniquely specified by
$\psi_d(x_i\eta)=\psi_d(x_i)\circ \psi_d(\eta)$ with
$\psi_d(x_i)(e)=x_0(d_i\shuffle e)$,
$i=0,1,\dots,m$
for any $e\in\allseries$,
and where $d_i$ is the $i$-th component series of~$d$
($d_0:=\mbf{1}$). By definition,
$\psi_d(\varnothing)$ is the identity map on $\allseries$.
A generalized series $\delta$ is defined as the unit for the composition product, that is,
$\delta\circ c=c\circ \delta=c$ for all $c\in\allseries$, or, equivalently, $F_\delta$ is the
identity map on the input space so that $F_\delta[u]=u$ for all admissible inputs $u$.

Finally, consider
the additive feedback connection with $F_c$ in the forward path and $F_d$ in the feedback path.
The generating series $e\in\allseries$ defines a closed-loop system $y=F_e[u]$ when it satisfies the feedback
equation $F_e[u]=F_c[u+F_{d\circ e}[u]]$.
It is known that there always exists such an $e=c@d$, where the series $c@d$ is the {\em feedback product} of $c$ and~$d$~\cite{Gray-etal_SCL14}.
This product has
an explicit formula in terms of the antipode of a graded connected Hopf algebra~\cite{Foissy_15}. Its construction relies in part on the
shuffle algebra.

\section{Entropy of generating series}\label{sec:entropy-of-generating-series}

Consider $\allseries$ as a graded $\re$-vector space with finite dimensional homogeneous components~$V_n$
spanned by all the words of degree~$n$, $X(n)$, so that $\dim(V_n)=|X(n)|$.
Given a series $c\in\allseries$ and $n\in \nat_0$, let $\supp_n(c):=\supp(c)\bigcap X(n)$.
Define the {\em support sequence} of $c$ to be $S_c=\{n_k\in \nat_0\colon |\supp_{n_k}(c)|\neq 0,\;k\in \nat_0\}$.
It is assumed that $|X(n)|\sim K\gamma^n$ for some real numbers $K,\gamma>0$ when $n\gg 0$.

\begde \label{de:entropy-generating-series}
The {\bf entropy} of a series $c\in\allseries/\allpoly$
with support sequence $S_c$
is
\[
h(c)=\limsup_{k\rightarrow \infty} \frac{1}{n_k}\log_{\gamma}(|\supp_{n_k}(c)|).
\]
If $c\in\allpoly$, then $h(c):=0$.
\endde

The entropy of a generating series is roughly a (normalized) measure of the rate at which its support
grows with increasing degree relative to the maximum rate it could grow as determined by the grading.

\begth \label{th:series-entropy-well-defined}
For every $c\in\allseries$, $h(c)$ is well defined with $0\leq h(c)\leq 1$.
\endth

\begin{proof}
The only nontrivial case is when $c\in\allseries/\allpoly$.
Consider the sequence of real numbers $a_k=\log_\gamma(|\supp_{n_k}(c)|)/n_k$, $n_k\in S(c)$, $k\geq 0$.
It is clearly bounded from below by zero. It is also bounded from above by one since no series in $\allseries$ has a larger support than
$\charseries(X^\ast)=\sum_{\eta\in X^\ast}\eta$, and
$\supp_k(\charseries(X^\ast))\sim K\gamma^k$, $k\gg 0$. As the infinite sequence $a_k$, $k\geq 0$ in $\re$ is bounded, it must have at least one cluster point.
The largest cluster point uniquely defines~$h(c)$.
\end{proof}

If the sequence $a_k$, $k\geq 0$ defined above converges, then clearly $h(c)$ is equivalent to this limit. In addition, it is evident that $h$ is
not homogeneous as $h(\alpha c)=h(c)$ for all nonzero $\alpha\in\re$.
Unless stated otherwise, the default assumption in most examples will be that $X=\{x_0,x_1\}$, $\allseries$ is graded by word length so that $K=1$,
$\gamma=2$, and either $S(c)=\nat_0$ or $S(c)=\nat$.

\begex \label{ex:one-letter-FPS-entropy}
A series (or polynomial) is called a {\em repeated word series} if
the only words in its support are powers of a single word $\xi\in X^+$. For example,
$c=x_1+x_1^2$ and $d=\sum_{n\geq 0} (x_0x_1x_0)^n$ are repeated word series.
Clearly, $h(c)=0$. In the latter case, where $\xi=x_0x_1x_0$, observe
$S_d=\{k|\xi|\colon k\geq 0\}$ and
$\supp_{k|\xi|}(d)=1$, $k\geq 0$, so that $h(d)=0$.
Therefore, all repeated word series have zero entropy.
\endex

\begex \label{ex:entropy-LTI-words}
A series $c\in\allseries$ is said to be {\em linear} if $|\eta|_{x_1}=1$ for all $\eta\in\supp(c)$.
For example, $c=\sum_{n\geq r} x_0^{n-1}x_1$ is a linear series with relative degree $r\geq 1$ \cite{Gray-Venkatesh_SCL2019}.
When $r=1$, for example, the input-output map $F_c$ is realized by the linear time-invariant system
\[
\dot{z}=z+u,\qquad z(0)=0,\qquad y=z.
\]
Observe $\supp_k(c)=1$, $k\geq r$ so that $h(c)=0$.
\endex

\begex
Consider the linear series $c=\sum_{n_0,n_1\geq 0} x_0^{n_0}x_1x_0^{n_1}$.
In this case, $\supp_k(c)=k+1$, $k\geq 0$, and thus, $h(c)=\lim_{k\rightarrow \infty} \log_2(k+1)/k=0$.
This implies that the generating series for any linear operator $F_c$ has zero entropy.
\endex

\begex \label{ex:entropy-x1-partitioned-words}
A series $c\in\allseries$ is said to be {\em input-limited} if for some fixed $N\in\nat$, $|\eta|_{x_1}\leq N$ for
all $\eta\in\supp(c)$. It is known in general that rationality is not preserved under the composition product.
However, $c\circ d$ is rational if $c$ and $d$ are rational, and $c$ is input-limited \cite{Ferfera_79,Ferfera_80}.
Consider, for example, the input-limited series
\[
c_N:=\sum_{n_0,n_1,\dots,n_N=0}^\infty x_0^{n_0}x_1x_0^{n_1}x_1\cdots x_0^{n_{N-1}}x_1x_0^{n_N}.
\]
The number of compositions (order partitions) of a nonnegative integer $K=n_0+n_1+n_2+\cdots+n_N$ into
$N+1$ parts where $n_i\geq 0$ is ${K+N\choose N}$.
Therefore,
$\supp_k(c)={k\choose N}$, $k\geq N$. Hence,
\begin{gather*}
h(c)=\lim_{k\rightarrow \infty} \frac{1}{k}\log_2 \left(\frac{k(k-1)\cdots(k-N+1)}{N!}\right)
\leq \lim_{k\rightarrow \infty} \frac{N}{k}\log_2 (k)-\frac{1}{k}\log_2 (N!)
=0,
\end{gather*}
so that $h(c)=0$. This is in fact the case for all input-limited series.
\endex

\begex \label{ex:palindromes}
Given a word $\xi\in X^\ast$, let $\tilde{\xi}$ denote the word whose letters are written in reverse order. For example, if $\xi=x_0x_1x_1$, then
$\tilde{\xi}=x_1x_1x_0$. Consider the set of all even {\em palindromes}
\[
P=\big\{\xi\tilde{\xi}\colon \xi\in X^\ast\big\}.
\]
Observe if $\eta\in P$ then $\tilde{\eta}\in P$. It can be directly checked that $\big|P\bigcap X^{2n}\big|=2^n$, $n\geq 0$. Setting $c=\sum_{\eta\in P} \eta$, it
follows that $S_{c}=\{2k\colon k\geq 0\}$ and $\supp_{2k}(c)=2^k$, $k\geq 0$. In which case, $h(c)=1/2$.
\endex

\begex \label{ex:word-power-series}
A {\em word power series} has the form $c_{1/N}=\sum_{\eta\in X^\ast} \eta^N$, where $\eta^N$ denotes the Cauchy product power and $N\in\nat$ is fixed.
Analysis similar to that in the previous example gives $h(c)=1/N$.
\endex

\begex
The word power series $\charseries(X^\ast)=\sum_{\eta\in X^\ast}\eta$ appearing in the proof of Theorem~\ref{th:series-entropy-well-defined} has entropy one
in light of the previous example, as
does the series $d=c-x_0^\ast$, where $x_0^\ast:=\sum_{k\geq 0} x_0^k$.
\endex

\begex \label{ex:bilinear-system-entropy-amplifier}
Consider the bilinear state space realization
\begdi
\dot{z}=z+zu,\qquad z(0)=1,\qquad y=z.
\enddi
Clearly, $\dim({\mathscr C}(X))=1$ since $g_0=g_1$ and $[g_0,g_1]=0$.
As an algebra, ${\mathscr C}(X)$ has zero entropy~\cite{Newman-etal_00}. But observe that
$(c,\eta)=L_\eta \output(z_0)=1$ for all $\eta\in X^\ast$. Thus, $c=\sum_{\eta\in X^\ast} \eta$ so that $h(c)=1$.
That is, the generating series of the input-output map has maximum entropy even though the
underlying state space realization has zero entropy in its control Lie algebra. This is possible in light of \rref{eq:FPS-differential-representation}
and the fact that $L_{[g_i,g_j]}h=L_{g_i}L_{g_j}h-L_{g_j}L_{g_i}h=0$ does not imply
that $L_{g_i}L_{g_j}h=L_{g_j}L_{g_i}h=0$.
This will be clearer in the next section when this example is revisited using
the shuffle algebra, which is playing a hidden role in this problem.
\endex

For any letter $x_i\in X$, let $x_i^{-1}$ denote the $\re$-linear {\em left-shift operator} defined by $x_i^{-1}(\xi)=\xi^\prime$
when $\xi=x_i\xi^\prime$ and zero otherwise.
It is defined
inductively for higher order shifts via $(x_i\xi)^{-1}=\xi^{-1}x_i^{-1}$,
where $\xi\in X^\ast$. The left-shift operator acts as a derivation on the shuffle product.
The {\em left-augmentation} of $c\in\allseries $ by $\xi\in X^\ast$ is simply the Cauchy product
$\xi c=\sum_{\eta\in X^\ast} (c,\eta) \xi\eta=\sum_{\eta\in X^\ast} (\xi c,\eta) \eta$.
Likewise, the {\em right-augmentation} of $c\in\allseries $ by $\xi\in X^\ast$ is
$c\xi=\sum_{\eta\in X^\ast} (c,\eta) \eta\xi=\sum_{\eta\in X^\ast} (c\xi,\eta) \eta$.

\begth \label{th:shifts-and-entropy}
For any series $c\in\allseries$ and word $\xi\in X^\ast$, $h\big(\xi^{-1}(c)\big)\leq h(c)$ and $h(\xi c)=h(c\xi)=h(c)$.
\endth

\begin{proof}
The first claim is a consequence of the fact that $\supp_k\big(\xi^{-1}(c)\big)\subseteq \supp_k(c)$, $k\geq 0$.
The equalities follow directly from the identities $|\supp_{k+\deg(\xi)}(\xi c)|=|\supp_{k+\deg(\xi)}(c\xi)|=|\supp_{k}(c)|$, $k\geq 0$.
\end{proof}

In general, entropy is invariant only under augmentations.

\begex
Suppose $c=\sum_{\eta\in X^\ast} x_0\eta$. Then $h(c)=1$, $h\big(x_0^{-1}(c)\big)=1$, and $h\big(x_1^{-1}(c)\big)=0$.
On the other hand, $h(x_ic)=h(cx_i)=1$, $i=0,1$.
\endex

\section{Entropy and interconnected nonlinear systems}\label{sec:entropy-interconnections}

\begde
A magma $(\allseries,\Box)$ with entropy function $h$ is said to be \textit{entropy bounded} if
\[
h(c\Box d)\leq \max(h(c),h(d)),\qquad \forall c,d\in\allseries.
\]
\endde

In essence, an entropy bounded product $\Box$ cannot create more entropy than that which is already present in its arguments.
The following is the first of two main results in this section.

\begth \label{th:entropy-bounded-products}
Addition, the Hadamard product, and the Cauchy product on $\allseries$ are all entropy bounded.
\endth

\begin{proof}
When necessary, it is sufficient to consider only the case where $c$ and $d$ are {\em positively supported}.
That is, if $\eta\in\supp(c)$, then $(c,\eta)>0$ and likewise for $d$.
For the products under consideration, the restriction to positively supported series will eliminate the
possibility of inter-term cancellations. Such cancellations can reduce the cardinality of the support of the
resulting series but never increase it. Without loss of generality, it is assumed below that all the support sequences are
equivalent to $\nat_0$.

{\em Addition.}
A slightly stronger result is possible when $c$ and $d$ are positively supported, namely, $h(c+d)=\max(h(c),h(d))$.
Otherwise, the possibility of cancellations makes the right-hand side only an upper bound.
Suppose $c$ and $d$ are positively supported.
Observe then that $\supp(c)\subseteq \supp(c+d)$, $\supp(d)\subseteq \supp(c+d)$, and
\[
\supp(c+d)= \supp(c)\cup \supp (d).
\]
Therefore, $|\supp_k(c)|\leq |\supp_k(c+d)|$, $|\supp_k(d)|\leq |\supp_k(c+d)|$, and
\begin{gather} \label{eq:supp-sums}
|\supp_k(c+d)|\leq |\supp_k(c)|+|\supp_k(d)|,\qquad \forall k\geq 0.
\end{gather}
Since $\log_2$ is an increasing function, the first two inequalities yield
$\max(h(c),h(d))\leq h(c+d)$. Consider a sequence $a=\{a_k\in\re\colon a_k\geq 1,\,k\in\nat_0\}$ such that
$\lambda(a):=\limsup_{k\rightarrow \infty} \log_2(a_k)/k$ is a real number.
It can be verified for any two such sequences $a$ and $b$ that
\begeq \label{eq:lambda-identity}
\lambda(a+b)=\max(\lambda(a),\lambda(b)).
\endeq
Hence, combining \rref{eq:supp-sums} and \rref{eq:lambda-identity} gives $h(c+d)\leq \max(h(c),h(d))$, and the claim is proved.

{\em Hadamard product.} If $\eta\in\supp_k(c\odot d)$, then $\eta\in\supp_k(c)\cap\supp_k(d)$.
But since $\supp_k(c)\cap\supp_k(d)\subseteq\supp_k(c)$ and $\supp_k(c)\cap\supp_k(d)\subseteq\supp_k(d)$, it follows that
\[
|\supp_k(c\odot d)|\leq \min(|\supp_k(c)|,|\supp_k(d)|),\qquad \forall k\geq 0.
\]
Therefore, in general,
\[
h(c\odot d)\leq \min(h(c),h(d)).
\]

{\em Cauchy product.}
Assume that $c,d\in\allseries$ are positively supported.
Then $\nu\in\supp_k(cd)$ if and only if
\[
(cd,\nu)=\sum_{\eta,\xi\in X^{\ast} \atop \nu=\eta\xi} (c,\eta)(d,\xi),
\]
where for at least one pair of words $(\eta,\xi)$ with $|\eta|+|\xi|=k$, the product $(c,\eta)(d,\xi)\neq 0$.
Therefore,
\[
|\supp_k(cd)|\leq\sum_{j=0}^k |\supp_{k-j}(c)||\supp_{j}(d)|,\qquad \forall k\geq 0.
\]
Note that in general this upper bound is conservative since $\nu$ can be constructed by concatenation of differently sized words,
i.e., $\nu=x_1x_0x_1=(x_1)(x_0x_1)=(x_1x_0)(x_1)$.
Now define the sequences $a=\{|\supp_{k}(c)|:k\geq 0\}$ and
$b=\{|\supp_{k}(d)|\colon k\geq 0\}$. Observe for $k\geq 1$ that
\begin{gather*}
\log_2\left(\sum_{j=0}^k a_{k-j}b_j\right) \leq \max_{0\leq j\leq k}\log_2(k a_{k-j}b_j)
 =\log_2(k)+ \max_{0\leq j\leq k}\log_2(a_{k-j}b_j).
\end{gather*}
Therefore,
\begin{align*}
\limsup_{k\rightarrow\infty} \max_{0\leq j\leq k}\frac{1}{k}\log_2(a_{k-j}b_j)&=
\limsup_{k\rightarrow\infty} \max_{0\leq j\leq k} \frac{k-j}{k}\left(\frac{1}{k-j}\log_2(a_{k-j})\right)+\frac{j}{k}\left(\frac{1}{j}\log_2(b_j)\right)\\
&\leq \max(\lambda(a),\lambda(b))
\end{align*}
since $(k-j/k)+(j/k)=1$ and the terms $\log_2(a_{k-j})/(k-j)$ and $\log_2(b_j)/j$ are asymptotically bounded by $\lambda(a)$ and $\lambda(b)$, respectively.
Thus, $h(cd)\leq \max(h(c),h(d))$ as claimed.
\end{proof}

It is assumed in the following set of examples that $X=\{x_0,x_1\}$.

\begex \label{ex:entropy-input-limited-series-with-limit}
Define $c_0$ to be the series in Example~\ref{ex:one-letter-FPS-entropy} when $\xi=x_0$.
Let $c_N$ denote the series in Example~\ref{ex:entropy-x1-partitioned-words} when $N\geq 1$.
Theorem~\ref{th:entropy-bounded-products} implies that for any integer $M\geq 0$, the
input-limited series $d_M=\sum_{N=0}^M c_N$
has zero entropy.
Note, however, that this
property is
not true in the limit since $\lim_{M\rightarrow\infty}d_M=\charseries(X^\ast)$ and $h(\charseries(X^\ast))=1$.
This indicates that $h\colon \allseries\rightarrow [0,1]$ is not continuous
when convergence on $\allseries$ is defined in the
ultrametric sense~\cite{Berstel-Reutenauer_88}.
\endex

\begex
Reconsider the series of even palindromes $c=\sum_{\eta\in P} \eta$ in Example~\ref{ex:palindromes}, where $h(c)=1/2$.
Let $d=\sum_{k\geq 0}x_0^k$ so that $h(d)=0$.
Then $c\odot d=\sum_{k=0}^\infty x_0^{2k}$ so that $h(c\odot d)=0=\min(1/2,0)$ as expected.
\endex

\begex \label{ex:zero-entropy-Cauchy-products}
All the series in Examples~\ref{ex:one-letter-FPS-entropy}--\ref{ex:entropy-x1-partitioned-words} can all be
written as the Cauchy product of repeated words series. For example, the series in Example~\ref{ex:entropy-x1-partitioned-words}
can be written as $c_N=d_0^\ast x_1d_1^\ast x_1\cdots d_{N-1}^\ast x_1 d_N^\ast$, where $d_i^\ast=\sum_{k\geq 0} x_0^k$.
Thus, by Theorem~\ref{th:entropy-bounded-products}, every series in these examples must have
zero entropy as was demonstrated directly from the definition.
\endex

\begex
Reconsider the series of even palindromes $c=\sum_{\eta\in P} \eta$ in Example~\ref{ex:palindromes} where $h(c)=1/2$.
The support of series $cc$ now contains elements that are {\em not} palindromes, for example, $x_0^2x_1^2$.
Nevertheless, Theorem~\ref{th:entropy-bounded-products} requires that $h(cc)\leq 1/2$. This can be verified by
a direct calculation. Note that every nonempty word in $\supp_k(cc)$ has the
form $\eta\tilde{\eta}\xi\tilde{\xi}$, where $2|\eta|+2|\xi|=k=2k^\prime$ with $k^\prime:=i+j$, $i:=|\eta|\geq 1$, and $j:=|\xi|\geq 1$.
Observe that $k^\prime:=i+j\geq 2$ has exactly ${k^\prime-1\choose 1}=k^\prime-1$ compositions with
two parts $i$ and $j$.
Thus, there are $(k^\prime-1)2^{k^\prime}$ words in $\supp_k(cc)$. In which case,
$\log_2(|\supp_k(cc)|)/k=\log_2((k^\prime-1)2^{k^\prime})/2k^\prime=1/2+\log_2(k^\prime-1)/2k^\prime$ so that in fact $h(cc)=1/2$.
\endex

The second main result of this section states that the remaining products of interest on $\allseries$ are
{\rm not} entropy bounded.
The claim is demonstrated via counterexamples presented subsequently.

\begth
The shuffle product, composition product, and feedback product on $\allseries$ are not entropy bounded.
\endth

\begex \label{ex:shuffle-product-not-entropy-bounded}
This example demonstrates that the shuffle product is not entropy bounded. If $X=\{x_0,x_1\}$, then
$x_i^\ast=\sum_{k=0}^\infty x_i^k$ has zero entropy.
It is known that \cite{Duffaut_Espinosa_09, Duffaut-Espinosa-etal_09}
\[
\charseries\big(X^k\big)
= \sum_{r_0,r_1\geq 0 \atop r_0+r_1=k}
x_0^{r_0}\shuffle x_1^{r_1},\qquad k\geq 0.
\]
Therefore,
\begin{gather*}
\sum_{\eta\in X^\ast}\eta =\charseries(X^\ast)
=\sum_{k=0}^\infty \charseries\big(X^k\big)
 =\sum_{k=0}^\infty \sum_{r_0,r_1\geq 0 \atop r_0+r_1=k}
x_0^{r_0}\shuffle x_1^{r_1}
 =x_0^\ast\shuffle x_1^\ast.
\end{gather*}
That is,
\[
h(x_0^\ast\shuffle x_1^\ast)=1>0=\max\{h(x_0^\ast),h(x_1^\ast)\}.
\]
On the other hand, there are specific instances where the entropy bounded
property holds, for example,
$
x_i^\ast\shuffle x_i^\ast=(2x_i)^\ast
$
so that $h(x_i^\ast\shuffle x_i^\ast)=h((2x_i)^\ast)=0$ \cite{Gray_fps-book}.
\endex

\begex
This example illustrates that the composition product is not entropy bounded.
Let $c=x_1^\ast$ so that $h(c)=0$. It is shown in
\cite{Gray-etal_SCL09} that
\[
\big(c\circ c,x_0^{k_0}x_1^{k_1}\cdots x_0^{k_{l-1}}x_1^{k_l}\big)
=(k_0)^{k_1}(k_0+k_2)^{k_3} \cdots
(k_0+k_2+k_4+\cdots+k_{l-1})^{k_l}
\]
for all odd $l\geq 1$ and $k_i\geq 0$, $i=0,1,\dots,l$ (assume $0^0:=1$).
Therefore, $\supp(c\circ c)=X^\ast$ and $h(c\circ c)=1$.
\endex

\begex \label{ex:Devlin-polynomials}
It is shown in this example that the feedback product is not entropy bounded.
Assume $X=\{x_0,x_1\}$, and let $c=\sum_{k\geq 0} k!\,x_1^k$ so that $h(c)=0$.
Let $d=\delta$, the composition unit.
Formally, $\delta\cap X^\ast$ is empty so that $h(\delta):=0$.
Consider an additive unity feedback interconnection with $F_c$ in the forward path and $F_\delta=I$ in the feedback path.
This defines a feedback product $e=c@\delta$.
The generating series $c@\delta$ was computed explicitly in \cite{Ebrahimi-Fard-Gray_IMRN17} using Hopf algebra methods described
in \cite{Duffaut_Espinosa-etal_JA16, Gray-etal_SCL14} and found to have the form
$c@\delta=\sum_{n\geq 1} b_n$, where
the polynomial sequence $b_n\in\allpoly$, $n\geq 1$ satisfies the linear
recursion
\begin{gather} \label{eq:Devlin-recursion}
	b_n=(n-1)b_{n-1}x_1+(n-2)b_{n-2}x_0,\qquad n\geq 2
\end{gather}
with $b_0=0$ and $b_1=1$.
These are the {\em Devlin polynomials} that appear in a combinatorial characterization of periodic solutions
to the Abel equation in the context of Hilbert's 16th problem \cite{Devlin_1989,Devlin_1991}.
The first few polynomials are:
\begin{gather*}
b_1 =1, \\
b_2 =x_1, \\
b_3 =2x_1^2+x_0, \\
b_4 =6x_1^3 +3x_0x_1+2x_1x_0, \\
b_5 =24x_1^4+12x_0x_1^2+8x_1x_0x_1+6x_1^2x_0+3x_0^2, \\
b_6 =120x_1^5+60x_0x_1^3+40x_1x_0x_1^2+30x_1^2x_0x_1+24x_1^3x_0+15x_0^2x_1+12x_0x_1x_0+8x_1x_0^2.
\end{gather*}
It is not hard to see from \rref{eq:Devlin-recursion} that $\supp(c@\delta)=X^\ast$
so that if $\allseries$ is graded
by word length then $h(c@\delta)=1$.
It is shown in \cite[Theorem~4.1]{Devlin_1989} that for $n\geq 1$
\[
\supp(b_n)=\{\eta\in X^\ast\colon \deg(\eta)=2|\eta|_{x_0}+|\eta|_{x_1}+1=n\}.
\]
Therefore, under the alternative grading in Example~\ref{ex:graded-vector-space},
where $|X(n)|\sim \big(1/\sqrt{5}\big)\varphi^n$ with $\varphi=\big(1+\sqrt{5}\big)/2$ when $m=1$, it is
evident that $h(c@\delta)=1$. It is worth noting that this second grading is
also the grading for the polynomial algebra of coordinate functions that defines the output feedback Hopf algebra used to
compute the general feedback product $c@d$ as discussed in Section~\ref{subsec:system-interconnections}.
\endex

The final two examples provide some applications of these
results in the context of nonlinear control theory.

\begex
Let $c$ and $d$ be generating series for two linear time-invariant systems as described in Example~\ref{ex:entropy-LTI-words}.
As linearity is preserved under the parallel sum, composition, and feedback
interconnections,
the resulting generating series in each case must have zero entropy.
The parallel product connection does not preserve linearity but nevertheless
yields a zero entropy generating series. To see this, observe that every word in the
support of $x_0^ix_1\shuffle x_0^jx_1$ must have the form $\xi x_1$ with $\xi\in X^{i+j-1}$
having the letter $x_1$ in any position. While it is {\em not} true that $x_0^\ast x_1\shuffle x_0^\ast x_1=x_0^\ast x_1 x_0^\ast x_1$,
it does hold that
$\supp((x_0^\ast x_1)\shuffle(x_0^\ast x_1))=\supp(x_0^\ast x_1 x_0^\ast x_1)$, and the latter
series has zero entropy as discussed
in Example~\ref{ex:zero-entropy-Cauchy-products}.
\endex

\begex
Reconsider the state space system in Example~\ref{ex:bilinear-system-entropy-amplifier} after the
coordinate transformation $z=\exp(\bar z)$ has been applied, namely,
\[
\dot{\bar z}=1+u,\qquad \bar z(0)=0,\qquad y=\exp(\bar z).
\]
The generating series is known to be invariant under a change of coordinates~\cite{Isidori_95}.
Observe $\bar{z}=F_{x_0+x_1}[u]$ so that
\begin{gather*}
y =\exp(F_{x_0+x_1}[u])
 =\sum_{n=0}^\infty (F_{x_0+x_1}[u])^n\frac{1}{n!}
 =\sum_{n=0}^\infty (F_{(x_0+x_1)^{\shuffle n}}[u])\frac{1}{n!},
\end{gather*}
using the shuffle power identity $F_c^n[u]=F_{c^{\shuffle n}}[u]$.
Therefore, $y=F_c[u]$, where
\[
c=\sum_{n=0}^\infty (x_0+x_1)^{\shuffle n}\frac{1}{n!}
=\sum_{n=0}^\infty \charseries(X^n)
=\sum_{\eta\in X^\ast} \eta,
\]
where the identity $\charseries(X^n)=(\charseries (X))^{\shuffle n}/n!$ has been used above.
What is clear in this coordinate system is that the output function is largely
responsible for the maximal growth in the entropy of the generating series $c$ due to
the appearance of the shuffle product.
Observe the vector fields in the state equation satisfy
$\bar{g}_0=\bar{g}_1$ so that $x_0\sim x_1$
and $[x_0,x_1]=0$. Hence, the entropy of the control Lie algebra is still zero as in the original
coordinate system.
\endex

\section{Entropy ultrametric space}

Two series $c,d\in\allseries$
are said to be {\em entropy equivalent}, denoted by $c\sim_h d$, when $h(c-d)=0$.
Using this equivalence relation, let $A=\allseries/{\sim_h}$ denote the quotient space.
No notational distinction between elements
of $\allseries$ and $A$ will be made except that $\mbf{0}$ will denote the equivalence class of zero entropy series in~$A$,
and $c=d$ is understood to be $c\sim_h d$ when $c,d\in A$.

\begth
The vector space $A=(A,\dist_h)$ with $\dist_h(c,d):=h(c-d)$ for $c,d\in A$ is a~bounded ultrametric space.
\endth

\begin{proof}
The only nontrivial property is the ultrametric inequality. Observe
that for any ${c{,}d{,}e\!\in\! A}$:
\begin{gather*}
\dist_h(c,d) =h(c-d)
 =h((c-e)+(e-d)) \\
 \hphantom{\dist_h(c,d)}{}
 \leq \max(h(c-e),h(e-d))
 =\max(\dist_h(c,e),\dist_h(e,d)).
\end{gather*}
The boundedness property is clear.
\end{proof}

Note that $\dist_h$ is not induced by a semi-norm as it is not homogeneous, recall $\dist_h(\alpha c,\alpha d)=\dist_h(c,d)$ for all nonzero $\alpha\in\re$.

\begex
In the present context, the series in Example~\ref{ex:entropy-input-limited-series-with-limit} satisfy $d_M=\mbf{0}$, $M\geq 0$.
So as a sequence in $A$, it trivially converges to $\mbf{0}$ in $A$.
\endex

\begex
The sequence $c_{1/N}$, $N\geq 1$ in Example~\ref{ex:word-power-series} defines a nontrivial sequence in $A$.
Since $\lim_{N\rightarrow \infty}\dist_h(c_{1/N},1)=\lim_{N\rightarrow\infty}h\big(\sum_{\eta\in X^{+}}\eta^N\big)=\lim_{N\rightarrow \infty} 1/N=0$, this sequence also converges
to $\mbf{0}$ in $A$.
\endex

An immediate question is in what sense, if any, is $A$ complete?
It is known under various notions of completeness that contractions on ultrametric spaces
have a unique fixed point or some approximate fixed point \cite{Priess-Crampe-Ribenboim_11,Priess-Crampe-Ribenboim_13}.
But at present, this is an open problem.

Let $X$ and $Y$ be two arbitrary alphabets. Any $\re$-linear mapping
$\tau\colon \allseries\rightarrow\allseriesY$ is called a {\em
transduction}~\cite{Fliess_74a,Jacob_75,Kuich-Salomaa_86,Salomaa-Soittola_78}.
It is completely specified by
\begin{equation*}
\tau(\eta)=\sum\limits_{\xi\in
Y^*}(\tau(\eta),\xi)\xi,\qquad \forall \eta\in X^*,
\end{equation*}
if for each $c\in\allseries$, the sum $\sum_{\eta\in X^\ast}(\tau(\eta),\xi)(c,\eta)$
is finite for all $\xi\in Y^\ast$.
Otherwise, it is only partially defined on $\allseries$.
One can canonically associate with any transduction $\tau$ a series in
$\allseriesXxY$, namely
\begdi \label{eq:trans2} \nonumber
\hat{\tau} = \sum\limits_{\eta\in X^*}\eta\otimes\tau(\eta)
= \sum\limits_{\eta\in X^*,\;\xi\in Y^*}(\tau(\eta),\xi)\,\eta\otimes\xi.
\enddi
From $\hat{\tau}$ define a second transduction $\tau^\prime\colon \allseriesY\rightarrow\allseries$ via
\begin{equation*}
\tau'(\xi)=\sum\limits_{\eta\in X^*}(\tau(\eta),\xi)\eta,\qquad \forall \xi\in Y^*.
\end{equation*}
$\tau^\prime$ is called the {\em inverse} of $\tau$. A transduction $\tau$
is called rational if the series $\hat{\tau}$ is a rational
series in $\allseriesXxY$. In which case, every rational series in
$\allseries$ is mapped to a rational series in $\allseriesY$ (assuming it
is well-defined).

There are a number of interesting open problems regarding transductions and
entropy. Suppose $\allseries$ and $\allseriesY$ are $N_0$-graded vector spaces. Then $\allseries\otimes \allseriesY$ is also
an $N_0$-graded vector space with
\[
(\allseries\otimes \allseries)_n=\bigoplus_{n=i+j} (\allseries)_i \otimes (\allseriesY)_j.
\]
If, as assumed in Section~\ref{sec:entropy-of-generating-series}, $(\allseries)_i$ and $(\allseriesY)_j$
are spanned by all the words of degree $i$ and $j$, respectively, then
likewise for the element $(\allseries\otimes \allseries)_n$.
Therefore, a given $\hat{\tau}$ has a well defined notion of entropy in terms of its formal power series representation as an
element in $\allseriesXxY$. If $c\in\allseries$ and $d\in\allseriesY$, then the first open question is how
are $h(c)$ and $h(\tau(c))$ related, and likewise for $h(d)$ and $h(\tau^\prime(d))$?
(Here $h$ refers to the appropriate definition of entropy based on its argument.)
In the event that $X=Y$, there are corresponding mappings $\tau_A\colon A\rightarrow A$ and
$\tau^\prime_A\colon A\rightarrow A$
between entropy equivalences classes.
Such a mapping is said to be {\em strictly contracting} if
\[
\dist_h(\tau_A(c),\tau_A(d))< \dist_h(c,d),\qquad \forall c\neq d\in A.
\]
A second open question is what notion of completeness is available for $A$ that would render a fixed point, i.e., $\tau_A(c)=c$
and/or $\tau_A^\prime(d)=d$?

\section{Conclusions}

A definition of entropy was introduced for the generating series of a Chen--Fliess series. The concept is most closely related to the notion that appears in formal language theory. It was shown to be trivial in the case of linear systems, i.e., all linear systems have zero entropy.
Formal power series products induced or related to system interconnections were then classified in terms of their ability to increase entropy relative
to their arguments. In particular, the shuffle product and all the products that utilize it in their definition can increase entropy. This result will
likely have applications in future work regarding networks of Chen--Fliess series.
Finally, the paper concluded by introducing an entropy ultrametric space as a possible context for the future analysis of feedback systems.

\subsection*{Acknowledgement}

The author would like to thank the referees for suggesting some significant clarifications and improvements to this paper.

\pdfbookmark[1]{References}{ref}
\LastPageEnding

\end{document}